\documentclass[final,1p,times,authoryear]{elsarticle}
\usepackage{longtable}
\usepackage{graphicx}
\usepackage{subcaption} 
\usepackage{amsmath,amsfonts,amssymb}
\usepackage{wrapfig}
\usepackage{multirow}
\usepackage{epsfig}
\usepackage{psfrag}
\usepackage{xcolor}
\usepackage{mathrsfs}
\usepackage{stackengine}
\usepackage{float}
\usepackage{longtable}
\usepackage{tabularx}
\journal{New Astronomy}

\begin{document}

\begin{frontmatter}

\title{On coordinate frames relevant for pulsar physics}

\author[a]{Jyotijwal Debnath\corref{cor1}}
\cortext[cor1]{Corresponding Author}
\ead{jyotijwal.debnath@inaf.it}
\author[b,c]{Manjari Bagchi}

\address[a]{INAF-Osservatorio Astronomico di Cagliari, Via della Scienza 5, I-09047 Selargius, Italy}
\address[b]{The Institute of Mathematical Sciences, C. I. T. campus, Taramani, Chennai, 600113, India}
\address[c]{Homi Bhabha National Institute, Training School Complex, Anushakti Nagar, Mumbai 400094, India}

\begin{abstract}
	
Pulsars can be either isolated or binary (or even in triple) systems. Observational features of pulsars are used to probe various aspects of fundamental physics, including emission mechanism, gravitational physics, etc. Theoretical models of various phenomena need different coordinate frames. Often different physical processes affect each other and we need to use multiple coordinate frames and relations between those. This short review presents an extensive set of coordinate frames and relations between those. 

\end{abstract}


\end{frontmatter}

\section{Introduction}
\label{sec:intro}

The investigation of pulsars has significantly advanced our understanding of several important areas in astrophysics including the internal physics of neutron stars and the behavior of matter under extreme physical conditions \citep{basu2025}. As pulsars act as remarkably precise cosmic clocks, with some millisecond pulsars exhibiting timing stability comparable to that of atomic clocks, these are also useful for testing general relativity and other theories of gravity \citep{bt14}, studying the properties of the interstellar medium, constructing a relativistic deep-space positioning system \citep{bcm11}, and searching for low-frequency gravitational waves \citep{agazie24}.

Tests of gravity are possible through pulsar timing analysis, which involves accurately calculating the rotational phases of the pulsar by modeling the delays between successive Times of Arrival (ToAs) of the pulses. There are various types of delays that the signal from a pulsar experiences, regardless of whether the pulsar is isolated or part of a binary system, e.g., dispersion delay, the solar system  R\"omer delay, the solar system Shapiro delay, etc. The signals from binary pulsars experience some additional delays mostly due to the orbital motion and the extra curvature of the spacetime around the companion. Some of these binary specific delays are the binary R\"omer delay, the binary Shapiro delay, the binary Einstein delay, delays due to light-bending effects, etc \citep{lorimer04}. 

Nearly ten percent of the presently known radio pulsars are members of binary systems.  Binaries in which the companion of the pulsar is also a compact object, such as a white dwarf, another neutron star, or a stellar-mass black hole, provide unique opportunities to test theories of gravity, including general relativity, scalar-tensor theory, and tensor-vector-scalar theory \citep{Stairs2003}.  

In most of the analytical studies aiming to model the emission physics or the gravitational physics or the effect of gravitation on the pulsar signal, one needs to use more than one coordinate frames. In this short review, we provide details of some of such frames that are commonly used. In Section \ref{sec:allframes}, we define these frames and in Section \ref{transrelation}, we present the transformation relations between these frames. Most of these frames have been used recently in studies of the effect of the light bending phenomenon on the signal of a pulsar in a binary, due to the gravitational field of the companion (\citealp{DBB23}; \citealp{DB25}).

\section{Some important frames in pulsar physics}
\label{sec:allframes}

Various reference frames commonly used in pulsar physics can be broadly classified into two classes, e.g., (i) frames needed to study isolated as well as binary pulsars and (ii) frames needed to study phenomena specific to binary pulsars.

\subsection{Frames needed to study isolated as well as binary pulsars}

\begin{figure*}
	\includegraphics[width=150mm]{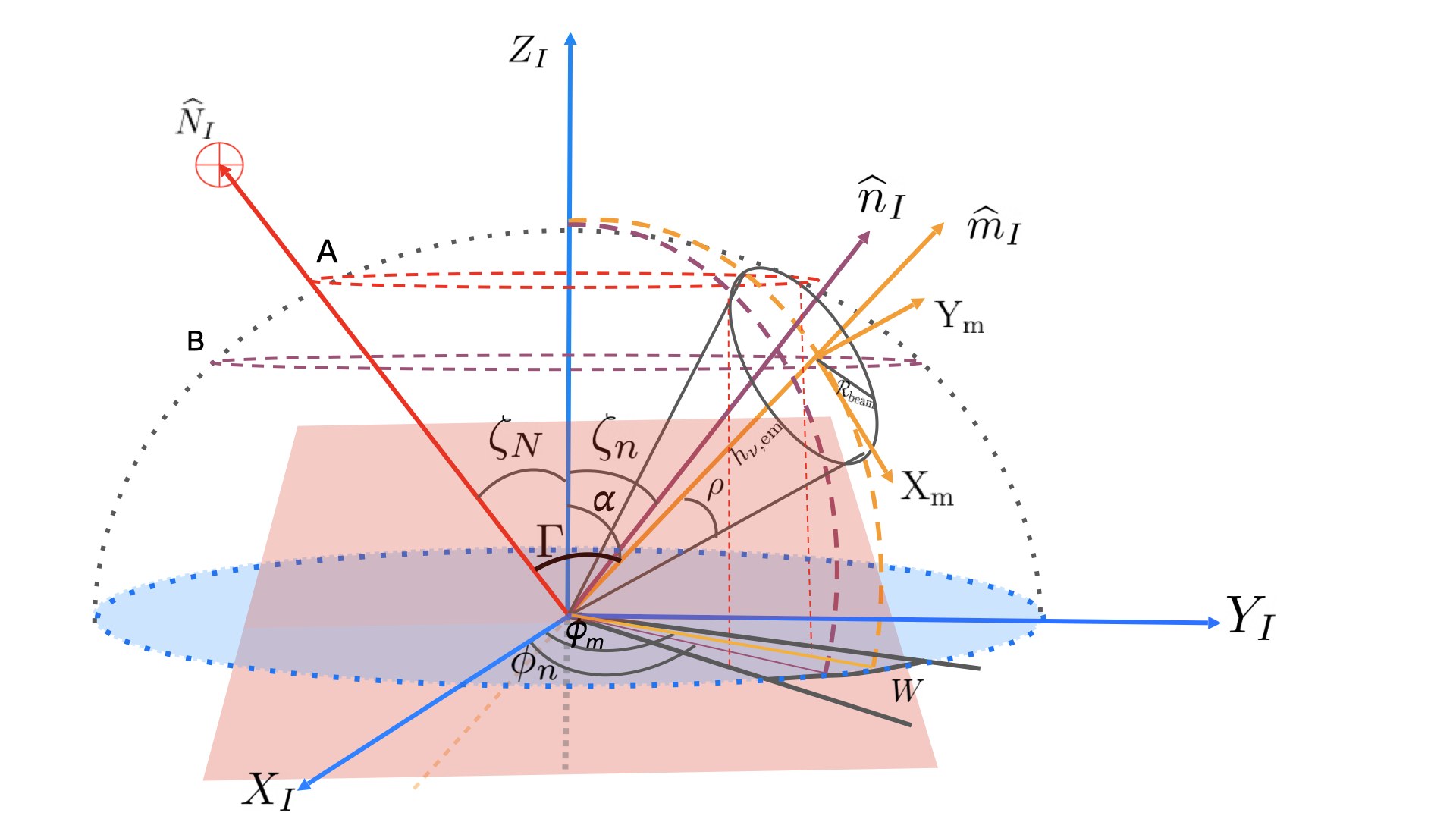}
	\caption{The beam of a pulsar and the Line-of-Sight (LoS). The meanings of various vectors and angles have been explained in the text. The red plane is perpendicular to the LoS, going through the pulsar, i.e., the sky-plane, and the blue plane is the $X_I Y_I$ plane. Note that here we have assumed that the pulsar rotates clockwise if viewed by the observer. If the case is reversed, then the $Z_I$-axis would be the negative direction of the spin axis. We always take $\widehat{m}_I$ along the side of the magnetic axis that is closer to the $Z_I$-axis.}
	\label{fig:spin_axis_frame}
\end{figure*}

Before discussing any frame, we first present the geometry of the pulsar beam. We know that a neuron star is viewed as a pulsar when its magnetic axis is misaligned from the spin axis and as the pulsar rotates, its magnetic axis also rotates around the spin axis \citep{rc69}. Fig. \ref{fig:spin_axis_frame} shows a schematic diagram of the emission geometry of a pulsar at an arbitrary time $t$. In this figure, we have shown different frames that are useful to study isolated pulsars, as we discuss next. We will define various angles while discussing these frames that are needed to understand the pulsar beam model in detail as discussed in Subsection \ref{subsubsect:beam_model}.

\subsubsection{The pulsar frame (I-frame)}
\label{subsubsect:I-frame} 
We define the `pulsar frame' (or the I-frame) with its origin at the center of the pulsar. The $Z$-axis of this frame (the $Z_I$-axis) is aligned with the spin axis of the pulsar. The line-of-sight (LoS) vector in this frame is defined as the direction from the center of the pulsar to the observer. The $X$-axis of the I-frame (the $X_I$-axis) is chosen to lie in the plane formed by the spin axis and the LoS vector. The unit vector along the LoS is denoted by $\widehat{N}_I$. 

The angle between the spin axis and the LoS is denoted by $\zeta_N$, so $\widehat{N}_I$ can be written as:   
\begin{equation}
\label{eq:N_I}
	\widehat{N}_I =
	\left[
	\sin \zeta_N,\,
	0,\,
	\cos \zeta_N
	\right].
\end{equation} The direction of the $Y$-axis of the I-frame (the $Y_I$-axis) is then determined by the right-handed cross product:  
\begin{equation}
\widehat{y}_I = \widehat{z}_I \times \widehat{x}_I.
\end{equation} where $\widehat{y}_I$, $\widehat{z}_I$, and $\widehat{x}_I$ are the unit vectors along the $X_I$, $Y_I$, and $Z_I$ axes, respectively.

Without any loss of generality, we also choose the magnetic axis to lie in the $X_I Z_I$ plane at time $t = 0$. Then, the rotational phase of the magnetic axis (or pulsar phase) becomes:  
\begin{equation}
	\label{pulsar_phase}
	\phi_m = \frac{2\pi}{P_s} t,
\end{equation}  
where $P_s$ is the spin period of the pulsar. Thus, $\phi_m$ is the angle between the $X_I$-axis and the projection of the magnetic axis onto the $X_I Y_I$ plane.  

The angle between the spin axis ($Z_I$) and the magnetic axis is denoted by $\alpha$. At any arbitrary time $t$, the unit vector along the magnetic axis $\widehat{m}_I$ can be written as:  
\begin{equation}
\label{eq:m_I}
	\widehat{m}_I = 
	\left[
	\sin \alpha \cos \phi_m,\,
	\sin \alpha \sin \phi_m,\,
	\cos \alpha
	\right].
\end{equation}

The angle between the magnetic axis and the LoS is $\Gamma$, which varies with the rotational phase ($\phi_m$) of the pulsar as the magnetic axis rotates about the spin axis. From Eqs. (\ref{eq:N_I}) and (\ref{eq:m_I}) we get,
\begin{equation}
\label{Gamma}
\Gamma=\cos ^{-1} \left( \sin \alpha \cos \phi_m \sin \zeta_N + \cos \alpha \cos \zeta_N \right )
\end{equation}
For a particular pulsar, among $\alpha$, $\phi_m$, and $\zeta_N $, only $\phi_m$ is time dependent. From Eq. (\ref{Gamma}), one can show that $\Gamma$ is minimum when $\phi_m=0$ and maximum when $\phi_m=\pi$. 
The minimum value of $\Gamma$ (at $\phi_m=0$), known as the angle of closest approach ($\beta$, not shown in the figure), can be written as:
\begin{equation}
\label{eq:angle_beta}
\beta = \zeta_N - \alpha.
\end{equation}

As shown in Fig. \ref{fig:spin_axis_frame}, $\widehat{n}_I$ is the unit vector along a generic light ray, which is considered to make an angle $\zeta_n$ with the $Z_I$-axis, while $\phi_n$ denotes the angle made by its projection onto the $X_I \, Y_I$ plane with the $X_I$-axis. Then,  $\widehat{n}_I$ can be written as:
\begin{equation}
	\widehat{n}_I =
	\left[
	\sin \zeta_n \cos \phi_n,\,
	\sin \zeta_n \sin \phi_n,\,
	\cos \zeta_n
	\right].
\end{equation}

This frame is particularly useful for studying various aspects of pulsar physics, such as the structure of the pulsar and its emission mechanism.

\subsubsection{The beam frame or magnetic axis frame (m-frame)}
\label{subsubsect:m-frame} 

We define the `beam frame' at the center of the beam cross-section, located at the position where the emission originates, i.e., at $h_{\nu, {  em}}$.  The $Z$-axis of the beam frame (the $Z_{m}$-axis) is aligned with the magnetic axis of the pulsar. The $X$ and the $Y$ axes of this frame (the $X_{ m}$ and the $Y_{ m}$ axes, respectively) span the plane perpendicular to $Z_m$, with the beam cross-section lying in the $X_{ m}  Y_{ m}$ plane. The $X_{  m}$-axis is chosen to lie in the plane formed by $Z_I$-axis and $Z_{ m}$-axis. The direction of the $Y_{  m}$-axis is then determined by the right-handed cross product:  
\begin{equation}
	\widehat{y}_m = \widehat{z}_m \times \widehat{x}_m, 
\end{equation} Here, $\widehat{x}_m$, $\widehat{y}_m$, and $\widehat{z}_m$ denote the unit vectors directed along the $X_m$, $Y_m$, and $Z_m$ axes, respectively. Hence, $\widehat{z}_m$ and $\widehat{m}_I$ are equivalent. This frame is particularly useful for studying properties associated with the emission mechanism, e.g., the beam geometry, internal structure of the emission cone, etc.

As shown in Fig. \ref{fig:spin_axis_frame}, as the beam rotates, the points of the beam that traverse the path marked `A' would fall along the LoS and would be visible. The magnetic axis traverses along the path `B' which is around the LoS and does not fall on it.

\subsubsection{Some parameters of a simple beam model}
\label{subsubsect:beam_model} 

We now try to understand a simple emission model with the help of various angles defined in earlier subsections. Many pulsar emission phenomena can be modelled with this simple emission model proposed by \citet{rc69}. According to this model, a pulsar emits radiation along a conal beam centered on its magnetic axis. The beam is generated at an emission height that depends on the observing frequency, given by \citet{lorimer04} as:
\begin{equation}
	\label{eq:emheight}
	h_{\nu, {  em}} = 400 \,\mathrm{km}
	\left( \frac{\nu}{10^{9}\,\mathrm{Hz}} \right)^{-0.26}
	\left( \frac{\dot{P}_{  s}}{10^{-15} \, {  s\,s^{-1}}} \right)^{0.07}
	\left( \frac{P_{  s}}{1\,\mathrm{s}} \right)^{0.30},
\end{equation}
where $\dot{P}_{  s}$ is the time derivative of the pulsar spin period ($P_{  s}$) and $\nu$ is the observing frequency.  

Following \citet{gks93}, the half-opening angle of the beam, $\rho$, is related to the observed pulse width $W$, the magnetic inclination $\alpha$, and the angle $\beta$, as given in Eq. (\ref{eq:angle_beta}) as:  
\begin{equation}
	\sin^2 \left(\frac{\rho}{2}\right)
	= \sin^2 \left(\frac{W}{4}\right)\sin \alpha \sin \zeta_N
	+ \sin^2 \left(\frac{\beta}{2}\right).
\end{equation}
Here, $W$ is the observed pulse width, which can be defined as:
\begin{equation}
W = \frac{\Delta t}{P_s} \times 2\pi,
\end{equation}
where $\Delta t$ is the total time the LoS remains inside the beam. In other words, $\Delta t$ is the total time during one rotation for which the angle between the magnetic axis and the LoS, i.e., $\Gamma$, remains less than the half-opening angle $\rho$ of the cone.

\subsection{Frames needed to study physics relevant for binary pulsars}

When a pulsar is in a binary system, it moves in an elliptical orbit around the barycenter (center of mass) of the system. In such a case, we often need to use some additional frames, some of these have the barycenter as the origin and some the center of the companion as the origin. We might even need to define frames centered at the center of the pulsar.

In case of a binary pulsar, the definition of LoS is different from that for an isolated pulsar. For a binary pulsar (or any binary stellar system) the LoS is defined from the barycenter to the observer \footnote{The definition for LoS for a single pulsar is different, conventionally, it is defined from the pulsar to the observer. However, the LoS for a single pulsar and the LoS for a binary pulsar are practically identical because, in any realistic scenario, the observer's distance is much larger than the orbital size. \label{footnote-LoS}}. 

The plane in which a pulsar and its companion move is known as the `orbital plane'. In Keplerian theory, this is a well defined plane. However, in the strong gravitational regime, when general relativistic effects are pronounced, the motions become more complicated, which is usually modeled as the change in the orientation of the orbital plane. 

\begin{figure*}
	\includegraphics[width=150mm]{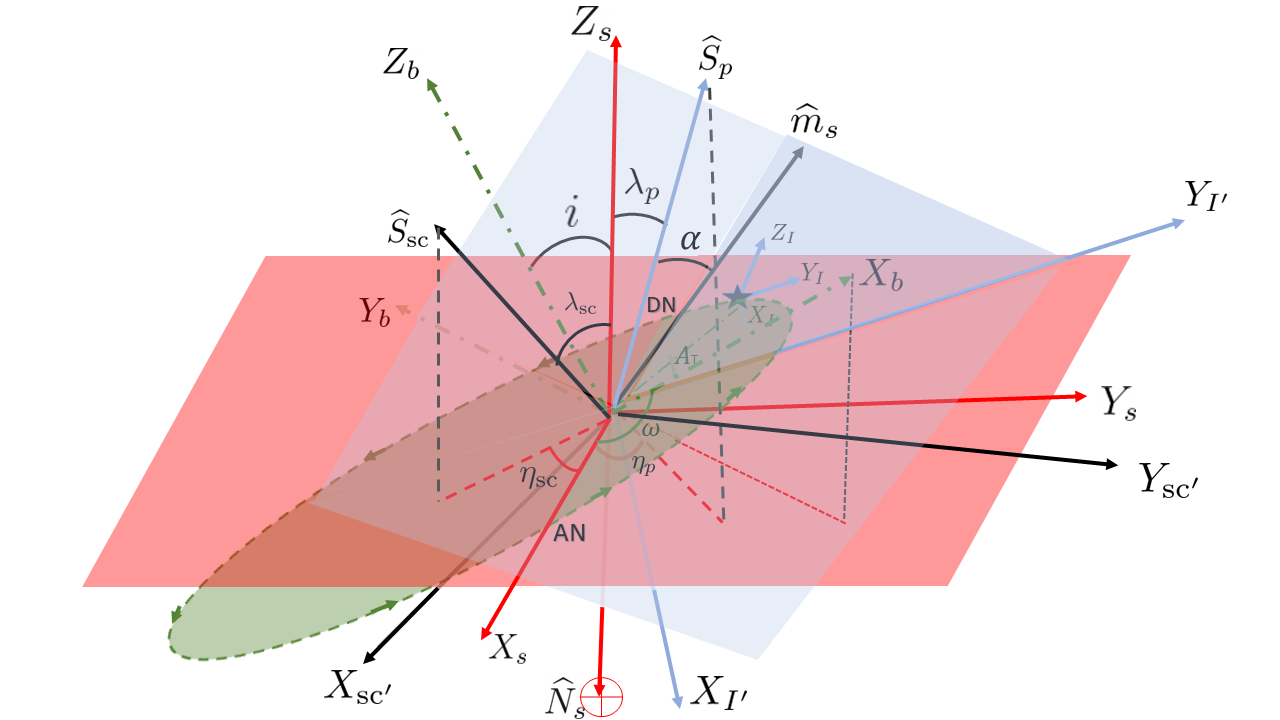}
	\caption{The orbital geometry of a pulsar (marked by a $\star$) in a binary system. The barycenter is taken as the origin of various frames shown. Details of the axes and angles can be found in the text. The sky plane (the $ X_s  Y_s$ plane) is shown with a light red color and the orbital plane (the $ X_b Y_b$ plane) is shown with a green color. Direction of motion of the pulsar on the orbital ellipse has been shown with green arrows on the ellipse. The light blue plane is the $X_I^{\prime}  Y_I^{\prime}$ plane that is parallel to the ${ X_I  Y_I}$ plane defined in Subsection \ref{I_to_s}. The unit vector along the spin axis of the pulsar is denoted by $\widehat{S}_p$, which is physically along the ${ Z_I}$-axis, but here shown along the ${ Z_I^{\prime}}$-axis, i.e., has been shifted parallely to the barycenter. Similarly, the unit  vector along the spin axis of the companion is denoted by $\widehat{S}_{sc}$, which is physically along the ${Z_I}$-axis, but here shown along the ${ Z_I^{\prime}}$-axis, i.e., has been shifted parallely to the barycenter. The ascending node has been denoted by AN and the descending node by DN. The direction of the earth is denoted by $\oplus$.}
	\label{fig:geometry}
\end{figure*}

One more important plane is the sky-plane'. In case of a binary pulsar, the sky-plane is the plane containing the barycenter and perpendicular to the LoS (shown as the red plane in Fig. \ref{fig:geometry}). For an isolated pulsar, the sky-plane can be taken simply as the plane perpendicular to the LoS and passing through the center of the pulsar (not shown in any of the figures).

\subsubsection{The sky frame (s-frame)}
\label{subsubsect:s-frame} 

To define the sky-frame or the s-frame, we first denote the unit vector along the LoS for a binary pulsar is $\widehat{N}_s$ (in this s-frame). The intersection of the sky-plane with the orbital plane of the binary is known as the `line of node'. The point of intersection of the orbital ellipse where the orbiting object (here the pulsar) moves away from the observer is called the `ascending node'(AN) and the other point of intersection where the orbiting object moves toward the observer is known as the `descending node' (DN).

Then, the s-frame is defined as the frame having origin at the barycenter and the $X$-axis and $Y$-axis are on the sky-plane, i.e., the sky plane can be called as the $X_s Y_s$ plane. The $Z$-axis of the s-frame (the $Z_s$-axis) is directed opposite to $\widehat{N}_s$. The $X$-axis ($X_s$) is chosen to lie along the line joining the barycenter and the ascending node (AN) of the orbit. The $Y$-axis ($Y_s$) is then determined using the right-handed cross product:  
\begin{equation}
	\widehat{y}_s = \widehat{z}_s \times \widehat{x}_s, 
\end{equation} where $\widehat{x}_s$, $\widehat{y}_s$, and $\widehat{z}_s$ denote the unit vectors directed along the $X_s$, $Y_s$, and $Z_s$ axes, respectively. This frame is often used in studies of orbital dynamics of pulsars and associated observational features.

\subsubsection{The sky frame shifted to the center of the pulsar (pS-frame) or to the center of the the companion (cS-frame)}
\label{subsubsect:pScS-frame}

Often, the vectors that are defined with respect to the center of the pulsas, need to be expressed in the sky-frame. For this purpose, one can simply shift the origin of the sky-frame to the center of the pulsar without changing the direction of the axes (translation without any rotation). This frame can be called the pS-frame whose $X$, $Y$, and $Z$ axes are denoted by $X_{pS}$, $Y_{pS}$, and $Z_{pS}$, respectively. 

Similarly, one can shift the origin of the sky-frame to the center of the companion without changing the direction of the axes (translation without any rotation). This frame can be called the cS-frame whose $X$, $Y$, and $Z$ axes are denoted by $X_{cS}$, $Y_{cS}$, and $Z_{cS}$, respectively. 

\citet{lorimer04} expressed the orientation of the spin axis (the ${Z_{I}}$-axis) of the pulsar in the pS-frame as follows. The angle between the ${ Z_{pS}}$-axis and the ${Z_{I}}$-axis is denoted by $\lambda_p$, and the angle between the $X_{pS}$-axis and the projection of the $Z_{I}$ axis on the $ X_{pS} \, Y_{pS}$-plane is denoted by $\eta_p$. Similarly, one can express the orientation of the spin axis of the companion in the cS-frame. The angle between the ${ Z_{cS}}$-axis and the ${Z_{sc}}$-axis is denoted by $\lambda_{sc}$, and the angle between the $X_{cS}$-axis and the projection of the $Z_{sc}$ axis on the $ X_{cS} \, Y_{cS}$-plane is denoted by $\eta_{sc}$. 

\begin{figure*}
	\includegraphics[width=150mm]{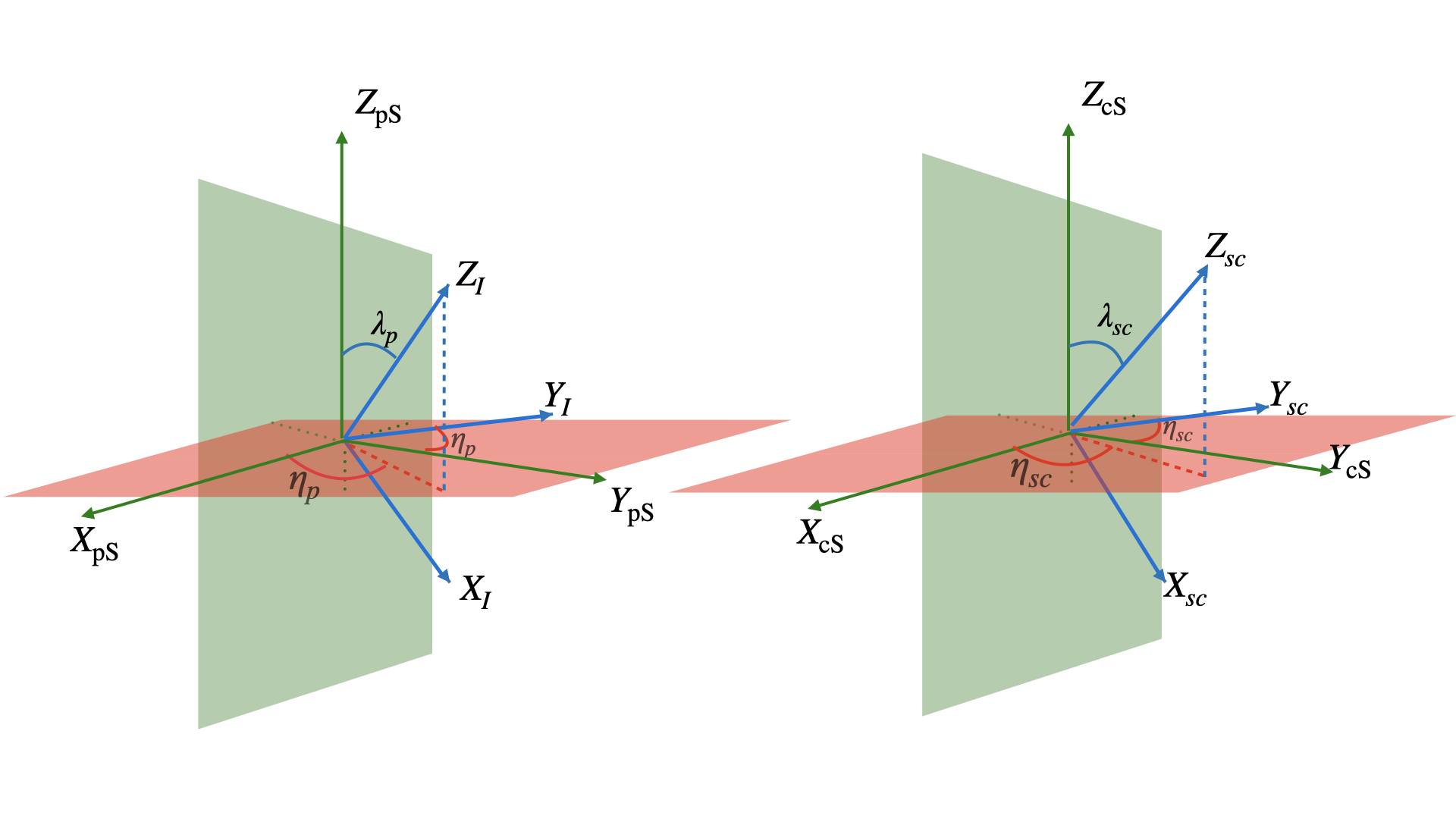}
	\caption{The left panel shows the I-frame and the pS-frame and the right panel shows the sc-frame and the cS-frame. These frames and various angles have been explained in the text.}
	\label{fig:pS_AND_cS}
\end{figure*}

In the left panel of Fig. \ref{fig:pS_AND_cS}, we show the orientation of the I-frame with respect to the pS-frame. As the pS-frame is parallel to the s-frame, I-frame is oriented exactly the same way with respect to the s-frame as depicted in Fig. \ref{fig:geometry}. Similarly, In the right panel of Fig. \ref{fig:pS_AND_cS}, we show the orientation of the sc-frame with respect to the cS-frame. As the cS-frame is parallel to the s-frame, sc-frame is oriented exactly the same way with respect to the s-frame as depicted in Fig. \ref{fig:geometry}.

\subsubsection{The orbital frame (b-frame)}
\label{subsubsect:b-frame}

We define the orbital frame with origin at the barycenter. The $Z_b$-axis is defined to be perpendicular to the orbital plane and pointed along the orbital angular momentum of the pulsar. In Fig. \ref{fig:geometry}, the orbital motion of the pulsar is clockwise if viewed from the observer, hence, the $Z_b$-axis and LoS are on the opposite side of the orbital plane. If the orbital motion was taken anti-clockwise if viewed from the observer, the $Z_b$-axis and LoS would be of the same side of the orbital plane. The $X_b$-axis lies along the line connecting the barycenter to the initial periastron of the orbit\footnote{Due to general relativistic effects, the orbital ellipse precesses in the orbit, one can model this periastron precession with respect to the $X_b$-axis.}.  Then, the $Y_b$-axis is then obtained from the cross product:  
\begin{equation}
	\widehat{y}_b = \widehat{z}_b \times \widehat{x}_b.
\end{equation} Here, $\widehat{x}_b$, $\widehat{y}_b$, and $\widehat{z}_b$ denote the unit vectors directed along the $X_b$, $Y_b$, and $Z_b$ axes, respectively.

This construction provides a well-defined right-handed coordinate system, referred to as the `orbital frame' (or the b-frame), which is used to study binary pulsar motion and to test predictions of general relativity.  

The inclination angle of the orbit is defined as the angle between ${Z_s}$-axis and ${ Z_b}$ and is denoted by $i$, and the angle between the ${ X_s}$-axis and the ${ X_b}$-axis is known as the longitude of the periastron and is denoted by $\omega$. These angles are also shown in Fig. \ref{fig:geometry}.

The above choice also implies that the angle between the the $ X_s Y_s$ plane and the the $ X_b Y_b$ plane, i.e., the angle between the sky-plane and the orbital plane is $i$, and this is often stated as the definition of the inclination angle.

\subsubsection{The spinning-companion frame or the sc-frame}
\label{subsubsect:sc-frame} 

It is evident that the companion in a binary pulsar system can itself be a spinning object, such as a rotating white dwarf, another pulsar, or a spinning black hole. In such cases, certain calculations, particularly geodesic solutions around the spinning companion, i.e., in Kerr spacetime, are most conveniently performed in a coordinate system centered on the spinning companion, named as the `spinning-companion frame' or the `sc-frame'. 

The $Z$-axis of this frame denoted by $Z_{\text{sc}}$, is chosen to align with the spin axis of the companion. The $X$-axis of the sc-frame (or the $X_{\text{sc}}$-axis), is chosen in such a way that the $Z_{\text{sc}} \, X_{\text{sc}}$ plane contains the LoS translated to the center of the companion. The direction of the $Y$-axis of the sc-frame (the $Y_{\text{sc}}$-axis) is then determined by the cross product
\begin{equation}
	\widehat{y}_{\text{sc}} = \widehat{z}_{\text{sc}} \times \widehat{x}_{\text{sc}} ,
\end{equation} where $\widehat{x}_{\text{sc}}$, $\widehat{y}_{\text{sc}}$, and $\widehat{z}_{\text{sc}}$ denote the unit vectors directed along the $X_{\text{sc}}$, $Y_{\text{sc}}$, and $Z_{\text{sc}}$ axes, respectively.

\subsubsection{The barycentric pulsar frame (I$^\prime$-frame)}
\label{subsubsect:Iprime-frame} 

The I-frame is defined with its origin at the center of the pulsar. We shift this origin to the barycenter keeping the directions of the axes unchanged. This gives a new frame called the I$^\prime$-frame with its $X$, $Y$, $Z$ axes labeled as $X_{I^{\prime}}$, $Y_{I^{\prime}}$, and $Z_{I^{\prime}}$, respectively. As $Z_I$ is taken along the spin axis of the pulsar, we can also say the spin vector of the pulsar is along the $Z_{I^{\prime}}$-axis.

As both the transformations, namely, the I$^\prime$-frame from the I-frame and the s-frame from the pS-frame (Subsection \ref{subsubsect:pScS-frame}) involve only translations without rotation, following Subsection \ref{subsubsect:pScS-frame}, we can say that the angle between the ${ Z_{S}}$-axis and the $Z_{I^{\prime}}$-axis is $\lambda_p$, and the angle between the $X_{s}$-axis and the projection of the $Z_{I^{\prime}}$ axis on the $ X_{s} \, Y_{s}$-plane is $\eta_p$. These angles have been shown in Fig. \ref{fig:geometry}.

\subsubsection{The barycentric spinning-companion frame or the sc$^\prime$-frame}
\label{subsubsect:scprime-frame} 

The sc-frame is defined with its origin at the center of the pulsar. We shift this origin to the barycenter keeping the directions of the axes unchanged. This gives a new frame called the sc$^\prime$-frame with its $X$, $Y$, $Z$ axes labeled as $X_{{sc}^{\prime}}$, $Y_{{sc}^{\prime}}$, and $Z_{{sc}^{\prime}}$, respectively. As the $Z_{sc^{\prime}}$-axis is taken along the spin axis of the companion, we can also say that the spin vector of the companion is along the $Z_{{sc}^{\prime}}$-axis. Hence, the unit vector along the $Z_{{sc}^{\prime}}$-axis , i.e, $\widehat{z}_{{sc}^{\prime}}$ is equivalent to the unit spin vector of the companion ($\widehat{S}_{sc}$) as shown in Fig. \ref{fig:geometry}.

As both the transformations, namely, the sc$^\prime$-frame from the sc-frame and the s-frame from the cS-frame (Subsection \ref{subsubsect:pScS-frame}) involve only translations without rotation, following Subsection \ref{subsubsect:pScS-frame}, we can say that the angle between the ${ Z_{s}}$-axis and the $Z_{sc^{\prime}}$-axis is $\lambda_{sc}$, and the angle between the $X_{s}$-axis and the projection of the $Z_{sc^{\prime}}$ axis on the $ X_{s} \, Y_{s}$-plane is by $\eta_{sc}$. These angles have been shown in Fig. \ref{fig:geometry}.

\subsubsection{The p-frame and the c-frame}
\label{subsubsect:p-frameANDc-frame} 

Two additional frames, not shown in Fig. \ref{fig:geometry}, are often useful. Those are defined at the centers of the pulsar and its companion: the $X_p Y_p Z_p$ frame (the `p-frame') and the $X_c Y_c Z_c$ frame (the `c-frame'), respectively.  

These frames are related to the b-frame through parallel translations along the line joining the pulsar, the barycenter, and the companion. 
Because the b-frame, the p-frame, and the c-frame are parallel to one another, any direction vector calculated in one frame will be identical in all three.  Moreover, the direction of the LoS is effectively the same in all three frames, as in any realistic scenario, the observer's distance is much larger than the orbital size.

\subsubsection{The T-frame}
\label{subsubsect:T-frame} 

In certain calculations involving binary pulsars, the initial direction of the emitted light rays plays a significant role. For example, the bending of light in the gravitational field of the companion depends sensitively on the initial propagation direction of the rays \citep{DBB23}. To handle such cases, we construct the `T-frame' for each of the light rays, depending on the initial direction of the light ray, but centered at the center of the companion (see Fig. \ref{fig:f3}).

The unit vector along the $Z$-axis of the T-frame or the $Z_T$-axis, denoted by $\widehat{z}_T$, is defined as:
\begin{equation}
	\label{eq:ZTb}
	\widehat{z}_T = \widehat{r}_b \times \widehat{n}_b ,
\end{equation}
where $\widehat{r}_b$ is the unit position vector of the center of the pulsar in the b-frame, and $\widehat{n}_b$ is the unit vector along the initial direction of the light ray from the pulsar (origin of the m-frame), also expressed in the orbital frame. The $Y$-axis of the T-frame is chosen to lie along the direction of $\widehat{r}_b$, i.e., the unit vector along the $Y_T$-axis is 
\begin{equation}
	\label{eq:YTb}
	\widehat{y}_T = \widehat{r}_b .
\end{equation} Finally, the unit vector along the $X$-axis of the T-frame ($ X_T$) is determined by the cross product of $\widehat{z}_T$ and $\widehat{y}_T$, 
\begin{equation}
	\label{eq:XTb}
	\widehat{x}_T = \widehat{z}_T \times \widehat{y}_T .
\end{equation} 

This frame is particularly useful for tracing null geodesics around a Kerr or Schwarzschild spacetime. We construct this frame such that the light ray initially lies in the $X_T Y_T$ plane. For Schwarzschild spacetime, the light ray does not change its plane of motion, meaning the $X_T Y_T$ plane remains the plane of the null geodesic throughout its propagation. In the Kerr spacetime, while the plane of motion can evolve over time, the light ray is still guaranteed to lie initially in the $X_T Y_T$ plane in both cases. 

Eq. (\ref{eq:YTb}) implies that the $Y_T$-axis lies on the $X_b Y_b$ plane (the orbital plane) and we define $\phi_{Tb}$ as the angle between the $Y_T$-axis and the $Y_b$-axis. We also denote the angle between the ${ Z_T}$-axis and the ${  Z_b}$-axis is denoted by $\theta_{ Tb}$. Hence, we can write, 
\begin{equation}
	\theta_{ Tb}=\cos^{-1}(\widehat{z}_{ T} \cdot \widehat{z}_{ b}) ,
	\label{eq:thetaTb}
\end{equation} and
\begin{equation}
	\phi_{ Tb} = \cos^{-1}(\widehat{y}_{ T} \cdot \widehat{y}_{b}) .
	\label{eq:phiTb}
\end{equation} 
The values of $\theta_{ Tb}$ and $\phi_{ Tb}$ can be calculated using Eqs. (\ref{eq:thetaTb}) and (\ref{eq:phiTb}) by using $\widehat{z}_T$ from Eq. (\ref{eq:ZTb})) and $\widehat{y}_T$ from Eq. (\ref{eq:YTb}). Moreover, as Eqs. (\ref{eq:XTb}), (\ref{eq:YTb}), and (\ref{eq:ZTb}) are in the b-frame, in Eqs.( \ref{eq:thetaTb}) and (\ref{eq:phiTb}), we can use $\widehat{z}_b=[0,0,1]$ and $\widehat{y}_b=[0,1,0]$.

\subsubsection{The L-frame}
\label{subsubsect:L-frame} 

\begin{figure}
	\centering
	\includegraphics[width=150mm]{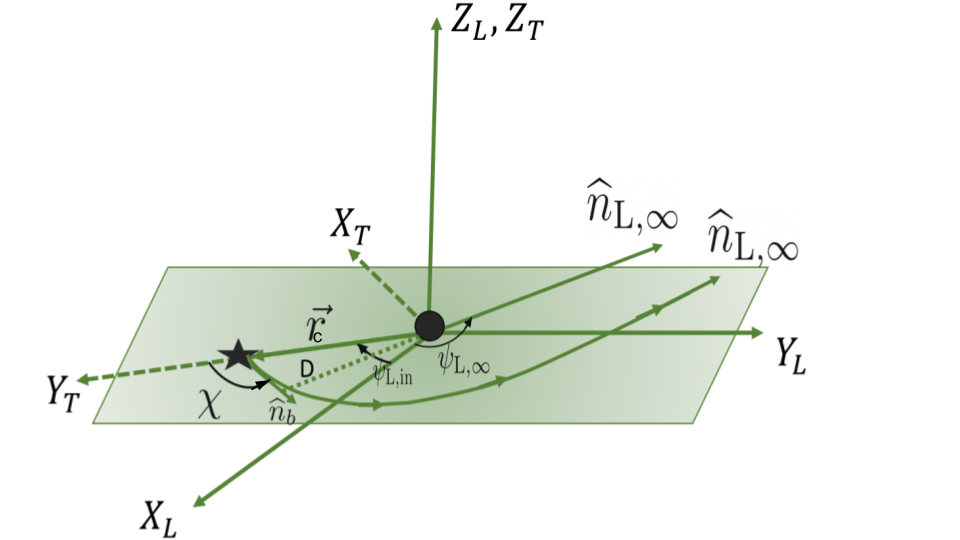}
	\caption{Schematic diagram illustrating the T-frame, the L-frame, and a light ray propagating toward a non-rotating gravitating companion ($\psi_{L,   in} < 0$). The light source (e.g., the pulsar) is indicated by a black star, and the companion is represented by a filled black circle. Coordinate axes and relevant angular parameters are explained in the text.}
	\label{fig:f3}
\end{figure}

\begin{figure}[h]
	\includegraphics[width=150mm]{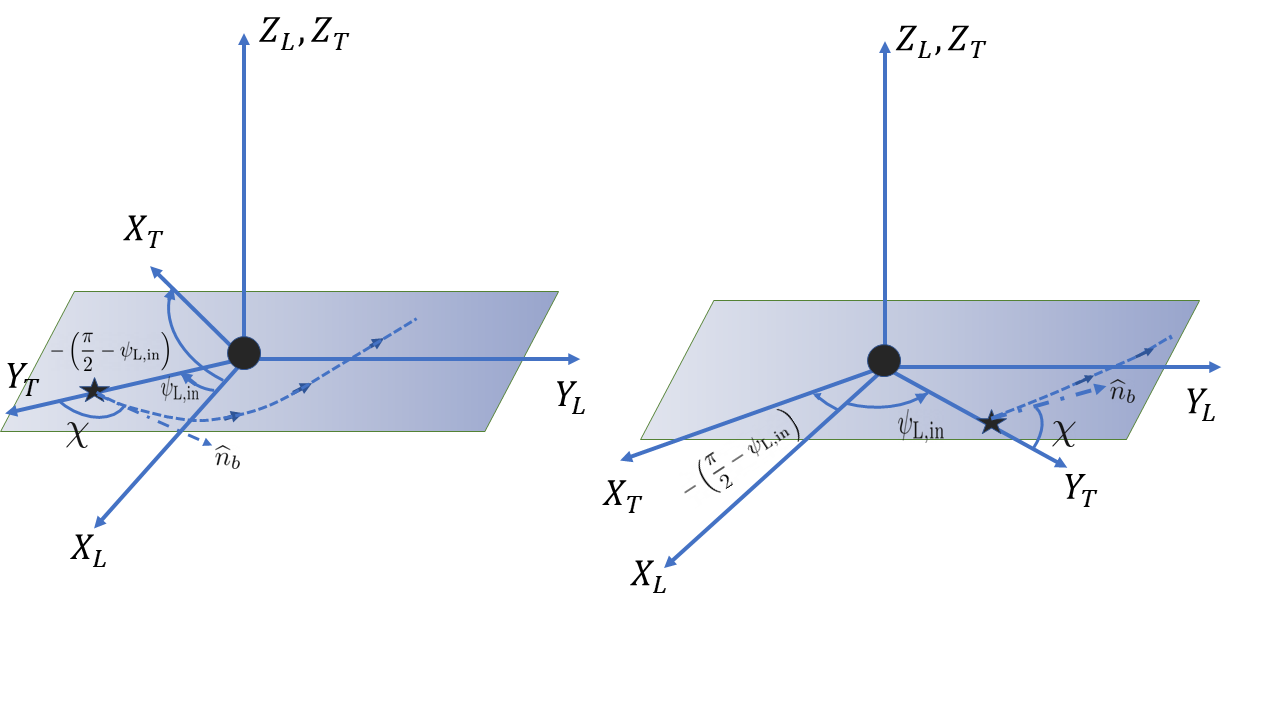}
	\caption{The orientation of the L-frame with respect to the T-frame. The blue plane is the $  X_L Y_L$ plane. The left figure is for the case when $\psi_{  L,in}$ is negative and the right figure is for the case when $\psi_{  L,in}$ is positive. In both of the panels, the pulsar is indicated by a black star, and the companion is represented by a filled black circle. }
	\label{fig:Transformation_from_L_to_T}
\end{figure}

In some analytical solutions of the null geodesic in Schwarzchild spacetime, as in \citet{chandrasekhar84}, one assumes that  the plane of the motion of a light ray (i.e, the $X_TY_T$ plane) is in the equatorial plane of the companion. Under this assumption, one does not lose generality, as such solutions are sought usually when the companion is also a compact object (another neutron star or a black hole) which is spherical when considered static. With such an assumption, we introduce an additional frame, namely the `L-frame' with origin at the center of the gravitating companion. This frame helps us solve the equations of the null geodesic in Schwarzchild spacetime following the formalism of \citet{chandrasekhar84}. We do not need the L-frame to solve the equations of the null geodesic in Kerr spacetime.

As mentioned earlier, in Schwarzschild spacetime, a light ray emitted from the pulsar travels in a plane defined by its initial propagation direction and containing both the pulsar and the gravitating companion. We take this plane as the $X_L Y_L$ plane of the L-frame, which is identical with the $X_T Y_T$ as defined in Subsection \ref{subsubsect:T-frame}. As the T-frame and the L-frame have the same origin, their $Z$-axis is also the same, i.e., $Z_T \equiv Z_L$. Hence, following Eq. (\ref{eq:ZTb}), the unit vector along the $Z_L$-axis (in the b-frame) can be written as:

\begin{equation}
	\label{eq:ZLb}
	\widehat{z}_L = \widehat{r}_b \times \widehat{n}_b ,
\end{equation}

Although the $X_L$-axis and the $Y_L$ axis lie in the $X_T Y_T$ plane, their directions are usually different from those of the $X_T$-axis and the $Y_T$-axis. For this, we denote the azimuthal angle of the light ray, i.e., the angle between the light ray and the $X_L$-axis as $\psi_L$.

The impact parameter $D$ of the light ray is defined as the perpendicular distance of its initial direction from the companion. If one knows the initial direction of the light ray and the position of the pulsar at the time of emission, then $D$ can be determined from the expression given by  \citet{Poutanen_2020}:  
\begin{equation}
\label{eq:impact_D}
	D = \frac{|\vec{r}_c| \sin \chi}{\sqrt{1 - \dfrac{GM_c}{|\vec{r}_c|^2}}},
\end{equation}
where $M_c$ is the mass of the companion, $G$ is the gravitational constant, $\chi$ is the angle between the pulsar's position vector in the c-frame, $\vec{r}_c$, and $\widehat{n}_b$ is the initial direction of the light ray in the b-frame, which is also the initial direction of the light ray in the c-frame (as the c-frame and the b-frame are parallel to each other). Fig. \ref{fig:f3} shows $D$, $\chi$, and $\widehat{n}_b$.

Using $D$ from Eq.(\ref{eq:impact_D}), and following the derivation of null geodesics from \citet{chandrasekhar84}, we can determine the initial azimuthal angle $\psi_{L,in}$ of the light ray in the $L$-frame, which is also the azimuthal angle of the source (e.g., the pulsar) in the L-frame at the time of emission. In the $T$-frame, we align the initial position of the source along the $Y_T$-axis. Hence, the angle between the $X_L$ axis and the $Y_T$ axis at the time of emission is $\psi_{L,in}$.

This makes the angle between the $X_L$-axis and the $X_T$-axis to be $|\pi/2 - \psi_{  L,in}|$. More specifically, we define the angle as $+ \left(\pi/2 - \psi_{  L,in}\right)$ when measured from the $X_T$-axis to the $X_L$-axis, and $- \left(\pi/2 - \psi_{  L,in}\right)$ when measured from the $X_L$-axis to the $X_T$-axis. As a result, a rotation of the $X_T$-axis by an angle $\left(\pi/2 - \psi_{  L,in}\right)$ about the $Z_T$-axis brings it into alignment with the $X_L$-axis (see Fig. \ref{fig:f3}). The $Y_L$-axis is then obtained using the following cross product:  
\begin{equation}
	\widehat{y}_L = \widehat{z}_L \times \widehat{x}_L .
\end{equation} where $\widehat{x}_L$, $\widehat{y}_L$, and $\widehat{z}_L$ are the unit vectors directed along the $X_L$, $Y_L$, and $Z_L$ axes, respectively.  

Note that, in the context of gravitational bending of the pulsar signal observed by a distant observer, only light rays with the impact parameter $D > 3\sqrt{3}GM_c/c^2$ are relevant \citep[c is the speed of light in vacuum]{chandrasekhar84}. When such a light ray reaches closest to the companion, it is taken as the $X_L$-axis, i.e, $\psi_L|_{  closest}=0$ (as shown in Fig. \ref{fig:f3}).

Now, as the requirement of our definition, we need to align the $Z$-axis of the L-frame, $Z_L$, with the $Z_T$-axis. To achieve this, we have adopted the convention that when the light ray initially moves towards the gravitating body, $\psi_{L, in}$ is negative, and when it initially moves away from the companion, $\psi_{L, in}$ is positive. These two situations are depicted in the left and the right panels of Fig. \ref{fig:Transformation_from_L_to_T}, respectively. 

The solution to the null geodesics yields the final azimuthal position of the light ray at infinity, denoted as $\psi_{L,\infty}$, which determines its asymptotic direction in the $L$-frame as 
\begin{equation}
    \widehat{n}_{L,\infty} = \begin{bmatrix} \cos \psi_{L,\infty} \\ \sin \psi_{L,\infty} \\ 0 \end{bmatrix}.
\end{equation}
We can then transform this unit vector into any relevant coordinate frame using the transformation relations presented in the next section.

\section{Transformation relations between different frames}
\label{transrelation}

In this section, we find the transformation relations between different frames and define the related angles to study the binary pulsar. To relate the frames already mentioned, sometimes we encounter some intermediate frames that we will see in due course. These intermediate frames are denoted with symbols prime ($\prime$) or double primes ($\prime \prime$). Two such primed frames, $I'$ and $\text{sc}'$, are illustrated in Fig.~\ref{fig:geometry}; these are related to their unprimed counterparts by a translation along the barycentric position vector. In addition, we introduce several intermediate coordinate frames that arise during the rotational transformations between these frames, as we discuss next.

In the discussion that follows, we use the symbol $R^P_{  N}(\theta)$ to denote a passive rotation about N-axis by an angle $\theta$ and $R^A_{  N}(\theta)$ to denote an active rotation about N-axis by an angle $\theta$, where N can be any of $X$, $Y$, $Z$.

\subsection{Transformation from the m-frame to the I-frame}
\label{m_to_I}
\begin{figure}[h]
	\centering
	\includegraphics[width=\linewidth]{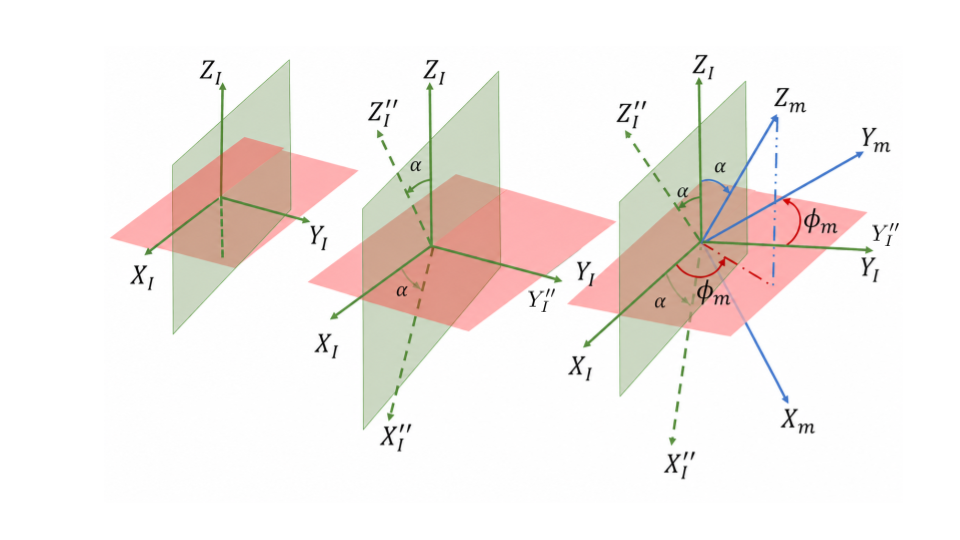}
	\caption{Transformation from the I-frame to the m-frame. The green plane is the $Z_I \, X_I$ plane, and the red plane is the $X_I \, Y_I$ plane.}
	\label{fig:Transformation_from_I_to_m}
\end{figure}

From the definitions described in Subsection \ref{subsubsect:I-frame}, we know that the angle between the $Z_I$-axis and the $Z_m$-axis is $\alpha$ and the projection of the $Z_m$-axis on the $X_I \, Y_I$-plane makes an angle $\phi_m$ with the $X_I$ axis. The transformation from the I-frame to the m-frame can be done in two steps as described below.

\begin{enumerate}[(i)]

\item By rotating the I-frame about the $  Y_I$-axis by an angle $\alpha$, we get a new frame ${  X^{\prime\prime}_I}{  Y^{\prime\prime}_I}{  Z^{\prime\prime}_I}$. Here $  Y^{\prime\prime}_I$ is along the $  Y_I$-axis. The $  Z^{\prime\prime}_I$-axis and the $  X^{\prime\prime}_I$-axis are in the $  Z_IX_I$ plane. The angle between the $  X_I$-axis and the $  X^{\prime\prime}_I$-axis is the same as the angle between the $  Z_I$-axis and the $  Z^{\prime\prime}_I$-axis, i.e., $\alpha$. This rotation is shown in the middle panel of Fig. \ref{fig:Transformation_from_I_to_m}.

\item Next, we rotate the ${  X^{\prime\prime}_I}{  Y^{\prime\prime}_I}{  Z^{\prime\prime}_I}$ frame about the $  Z_I$-axis by an angle $\phi_m$ to get the ${  X_m}{  Y_m}{  Z_m}$ frame. As the rotation is about the $  Z_I$-axis, $  Y_m$ remains in the $  X_I Y_I$ plane (the red plane in  Fig. \ref{fig:Transformation_from_I_to_m}). This rotation is shown in the right panel of Fig. \ref{fig:Transformation_from_I_to_m}. 

\end{enumerate}

Together, these transformations can be summarized as:  
\begin{equation}
	{  X_I \, Y_I \, Z_I} \xrightarrow[]{R^P_{  Y_I}(\alpha)}  {  X_I^{\prime\prime} \, Y_I^{\prime\prime} \, Z_I^{\prime\prime} } \xrightarrow[]{R^P_{  Z_I}(\phi_m)}  {  X_m \, Y_m \, Z_m} ~.
	\label{eq:Itrans1}
\end{equation} 
Hence, the reverse transformation can be written as:
\begin{equation}
	{  X_m \, Y_m \, Z_m} \xrightarrow[]{R^P_{  Z_I}(-\phi_m)}  {  X_I^{\prime\prime} \, Y_I^{\prime\prime} \, Z_I^{\prime\prime} } \xrightarrow[]{R^P_{  Y_I}(-\alpha)}  {  X_I \, Y_I \, Z_I} ~.
	\label{eq:Itrans2}
\end{equation}

Thus, if a vector is denoted by $\overrightarrow{V}_{  m}$ in the m-frame and by $\overrightarrow{V}_{  I}$ in the I-frame, these two would be related as 
\begin{equation}
	\overrightarrow{V}_{  I}  = R^A_{  Z}(\phi_m) \,  R^A_{  Y}(\alpha) \overrightarrow{V}_{  m} ~.
	\label{eq:mtransI}
\end{equation}

\subsection{Transformation from the I-frame to the I$^\prime$-frame}
\label{I_to_Iprime}

As described in Subsection \ref{subsubsect:Iprime-frame}, the I-frame (origin at the center of the companion) and the I$^\prime$-frame (origin at the center of the pulsar) are related by a translation without any rotation, any vector in these two frames is essentially the same, i.e., if a vector in the I-frame be denoted by $\vec{V}_{I}$ and the same vector in the I$^\prime$-frame by $\vec{V}_{I^\prime}$ we can write,  
\begin{equation}
	\vec{V}_{  I} = \vec{V}_{  I^\prime}.
\end{equation}

\subsection{Transformation from the I$^\prime$-frame to the s-frame}
\label{sub:Iprime_to_s}

In Subsection \ref{subsubsect:pScS-frame}, we discussed how the orientation of the spin axis of the pulsar, i.e., the $Z_I$-axis is specified with respect to the pS-frame. As the I$^\prime$-frame is obtained from the I-frame with a translation without any rotation (Subsection \ref{I_to_s}), the $Z_I$-axis and the $Z_I^{\prime}$-axis have the same orientation. Moreover, as the pS-frame is obtained from the s-frame with a translation without any rotation, the orientation of the $Z_I^{\prime}$-axis in the s-frame is the same as the orientation of the $Z_I$-axis in the pS-frame.

Hence, following Subsection \ref{subsubsect:pScS-frame}, the angle between the ${ Z_s}$-axis and the ${Z_{I^\prime}}$-axis is denoted by $\lambda_p$, and the angle between the $X_s$-axis and the projection of the ${  Z_{I^\prime}}$ axis on the ${ X_s \, Y_s}$-plane is denoted by $\eta_p$. Using this information, it is easier to visualize the transformation of the s-frame to the I$^\prime$-frame in two steps.

\begin{figure}[h]
	\includegraphics[width=150mm]{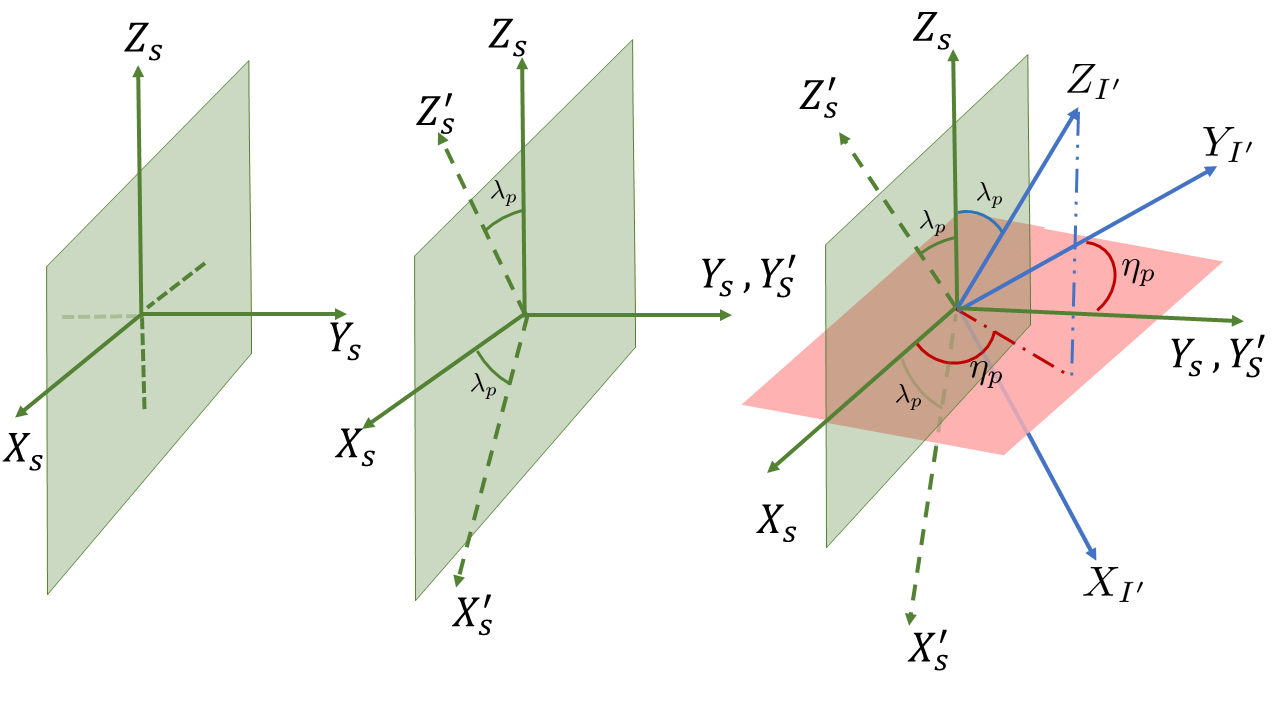}
	\caption{The transformation from the s-frame to the I-frame. The green plane is the $Z_s X_s$ plane and the red plane is the $X_s Y_s$ plane.}
	\label{fig:Transformation_from_S_to_I}
\end{figure}

\begin{enumerate}

\item A rotation of the s-frame about the ${ Y_s}$-axis by an angle $\lambda_p$ gives a new frame ${ X_s^{\prime} \, Y_s^{\prime} \, Z_s^{\prime}}$ where ${ Y_s^{\prime}}$ is along the ${ Y_s}$-axis, i.e., the ${Y_s^{\prime}}$-axis is in the ${ X_s \, Y_s}$ plane. The $Z_s^{\prime} $-axis and the $X_s^{\prime} $-axis are in the $Z_s \, X_s$ plane. The angle between the $Z_s $-axis and the $Z_s^{\prime} $-axis as well as the angle between the $X_s$-axis and the $X_s^{\prime} $-axis is $\lambda_p$. This transformation is shown in the middle panel of Fig. \ref{fig:Transformation_from_S_to_I}.

\item We rotate the ${ X_s^{\prime} \, Y_s^{\prime} \, Z_s^{\prime}}$ frame about the ${Z_s}$-axis by an angle $\eta_p$ to get ${ X_{I^\prime} \, Y_{I^\prime} \, Z_{I^\prime}}$ frame. As the rotation is about the ${  Z_s}$-axis, the ${Y_{I^\prime}}$-axis remains in the ${ X_s \, Y_s}$ plane. This transformation is shown in the middle panel of Fig. \ref{fig:Transformation_from_S_to_I}.

\end{enumerate}

Together, the above two-step transformation can be represented as:
\begin{equation}
	{ X_s \, Y_s \, Z_s} \xrightarrow[]{R^P_{ Y_s}(\lambda_p)}  {  X_s^{\prime} \, Y_s^{\prime} \, Z_s^{\prime} } \xrightarrow[]{R^P_{ Z_s}(\eta_p)}  { X_{I^\prime} \, Y_{I^\prime} \, Z_{I^\prime}} ~.
	\label{eq:ptrans1pr}
\end{equation}

Hence, the reverse transformation can be written as:
\begin{equation}
	{ X_{I^\prime} \, Y_{I^\prime} \, Z_{I^\prime}} \xrightarrow[]{R^P_{ Z_s}(-\eta_p)}  { X_s^{\prime} \, Y_s^{\prime} \, Z_s^{\prime} } \xrightarrow[]{R^P_{ Y_s}(-\lambda_p)}  { X_s \, Y_s \, Z_s} ~.
	\label{eq:ptrans2pr}
\end{equation}

Thus, if a vector is denoted by $\overrightarrow{V}_{I}$ in the I-frame and by $\overrightarrow{V}_{ s}$ in the s-frame, these two would be related by
\begin{equation}
	\overrightarrow{V}_{ s}  = R^A_{ Z}(\eta_p) \,  R^A_{ Y}(\lambda_p) \overrightarrow{V}_{ I^{\prime}} ~.
	\label{eq:atrans1pr}
\end{equation}  

\subsection{Transformation from the I-frame to the s-frame}
\label{I_to_s}

As we mentioned in Subsection \ref{I_to_Iprime}, the I-frame and the I$^\prime$-frame are identical, the above transformations relate the I-frame to the s-frame as:
\begin{equation}
	{ X_s \, Y_s \, Z_s} \xrightarrow[]{R^P_{ Y_s}(\lambda_p) \, R^P_{ Z_s}(\eta_p)}  { X_{I} \, Y_{I} \, Z_{I}} ~ ,
	\label{eq:ptrans1}
\end{equation} and the reverse transformation as
\begin{equation}
	{ X_{I} \, Y_{I} \, Z_{I}} \xrightarrow[]{R^P_{ Z_s}(-\eta_p) \, R^P_{ Y_s}(-\lambda_p)}    { X_s \, Y_s \, Z_s} ~.
	\label{eq:ptrans2}
\end{equation}

Thus, if a vector is denoted by $\overrightarrow{V}_{I}$ in the I-frame and by $\overrightarrow{V}_{ s}$ in the s-frame, these two would be related by
\begin{equation}
	\overrightarrow{V}_{ s}  = R^A_{ Z}(\eta_p) \,  R^A_{ Y}(\lambda_p) \overrightarrow{V}_{ I} ~.
	\label{eq:atrans1}
\end{equation}

\subsection{Transformation from the s-frame to the b-frame}
\label{s_to_b}

\begin{figure}[h]
	\includegraphics[width=150mm]{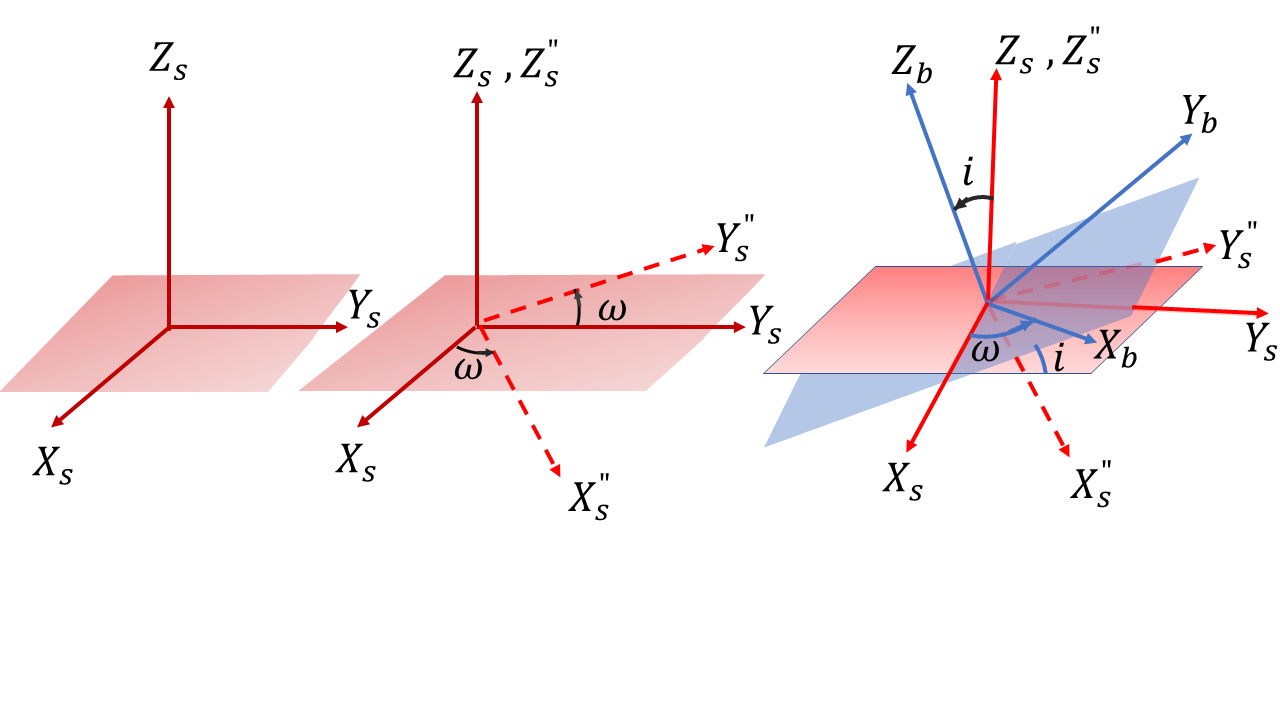}
	\caption{The transformation from the s-frame to the b-frame. The red plane is the $X_s Y_s$ plane (the sky plane), and the blue plane is the $ X_b Y_b$ plane (the orbital plane). The $ X_s$-axis is the line of the nodes, which is along the intersection of the sky-plane and the orbital plane.}
	\label{fig:Transformation_from_S_to_b}
\end{figure}

We have mentioned in Subsection \ref{subsubsect:b-frame}, that the angle between ${ Z_s}$-axis and ${ Z_b}$ is denoted by $i$ and the angle between the ${ X_s}$-axis and the ${ X_b}$-axis is denoted by $\omega$. Hence, the b-frame can be obtained from the s-frame through the following two steps:

\begin{enumerate}

\item A rotation of the s-frame about the ${ Z_s}$-axis by an angle $\omega$ gives a new frame ${ X_s^{\prime \prime} \, Y_s^{\prime \prime} \, Z_s^{\prime \prime}}$ where ${ Z_s^{\prime \prime}}$ is along the ${ Z_s}$-axis and the ${ X_s^{\prime \prime}}$-axis and the ${ Y_s^{\prime \prime}}$-axis are in the ${ X_s \, Y_s }$ plane. This transformation is shown in the middle panel of Fig. \ref{fig:Transformation_from_S_to_b}. The angle between the $X_s $-axis and the $X_s^{\prime \prime} $-axis as well as the angle between the $Y_s$-axis and the $Y_s^{\prime \prime}$-axis is $\omega$. 

\item Now, a rotation of the ${ X_s^{\prime \prime } \, Y_s^{\prime \prime} \, Z_s^{\prime \prime}}$ frame about the ${ X_s}$-axis by an angle $i$ gives the ${ X_b \, Y_b \, Z_b}$ frame. This transformation is shown in the right panel of Fig. \ref{fig:Transformation_from_S_to_b}, which ensures that the angle between the $ X_s^{\prime \prime}$-axis and the ${  X_b}$-axis, between the ${Y_s^{\prime \prime}}$-axis and the ${ Y_b}$-axis, as well as that between the ${ Z_s^{\prime \prime}}$-axis and the ${ Z_b}$-axis are $i$.

\end{enumerate}

Together, the above two-step transformation can be represented as:
\begin{equation}
	{ X_s \, Y_s \, Z_s} \xrightarrow[]{R^P_{ Z_s}(\omega)}  { X_s^{\prime \prime} \, Y_s^{\prime \prime} \, Z_s^{\prime \prime} } \xrightarrow[]{R^P_{ X_s}(i)}  { X_b \, Y_b \, Z_b} ~.
	\label{eq:ptrans3}
\end{equation}

Thus, if a vector is denoted by $\overrightarrow{V}_{ s}$ in the s-frame and by $\overrightarrow{V}_{ b}$ in the b-frame, these two would be related by
\begin{equation}
	\overrightarrow{V}_{ b}  = R^A_{ Z}(-\omega) R^A_{ X}(-i)   \overrightarrow{V}_{ s} ~.
	\label{eq:atrans2}
\end{equation} or,
\begin{equation}
	\overrightarrow{V}_{ s}  = R^A_{ X}(i)  R^A_{ Z}(\omega)  \overrightarrow{V}_{ b} ~.
	\label{eq:atrans_reverse2}
\end{equation}

Note that, instead of Eq. (\ref{eq:ptrans3}), if the second transformation is a rotation of the ${ X_s^{\prime \prime } \, Y_s^{\prime \prime} \, Z_s^{\prime \prime}}$ frame about the ${ Y_s}$-axis by an angle $i$, we will still get a new frame whose $ Z$-axis would make an angle $i$ with the $ Z_s$-axis, but the $  X_s$-axis will not remain along the intersection of the sky-plane (the ${ X_s \, Y_s}$ plane) and the ${ X \, Y}$ plane of the new frame, i.e., the $ X_s$-axis will not be the line of node, if we assume the ${ X \, Y}$ plane of the new frame as the b-frame. This situation is contrary to our definition of the $X_s$-axis being along the line of node.

\subsection{Transformation from the I-frame to the b-frame}

To express vectors in the b-frame, we proceed in two steps:
\begin{enumerate}
	\item Transform the vector from the I-frame to the s-frame as described in Subsection \ref{I_to_s}.
	\item Transform the result from the s-frame to the b-frame as described in Subsection \ref{s_to_b}.
\end{enumerate}

Combining Eqs. (\ref{eq:atrans2}) and (\ref{eq:atrans1pr}), we get,

\begin{equation}
	\overrightarrow{V}_{ b}  = R^A_{  Z}(-\omega) \, R^A_{ X}(-i)   \vec{V_s} ~ = R^A_{ Z}(-\omega)\, R^A_{ X}(-i)\, R^A_{ Z}(\eta_p) \,  R^A_{ Y}(\lambda_p) \, \overrightarrow{V}_{ I} ~.
	\label{eq:atrans3}
\end{equation}

\subsection{Transformation from the T-frame to the b-frame}
\label{T_to_b}

\begin{figure}[h]
	\includegraphics[width=150mm]{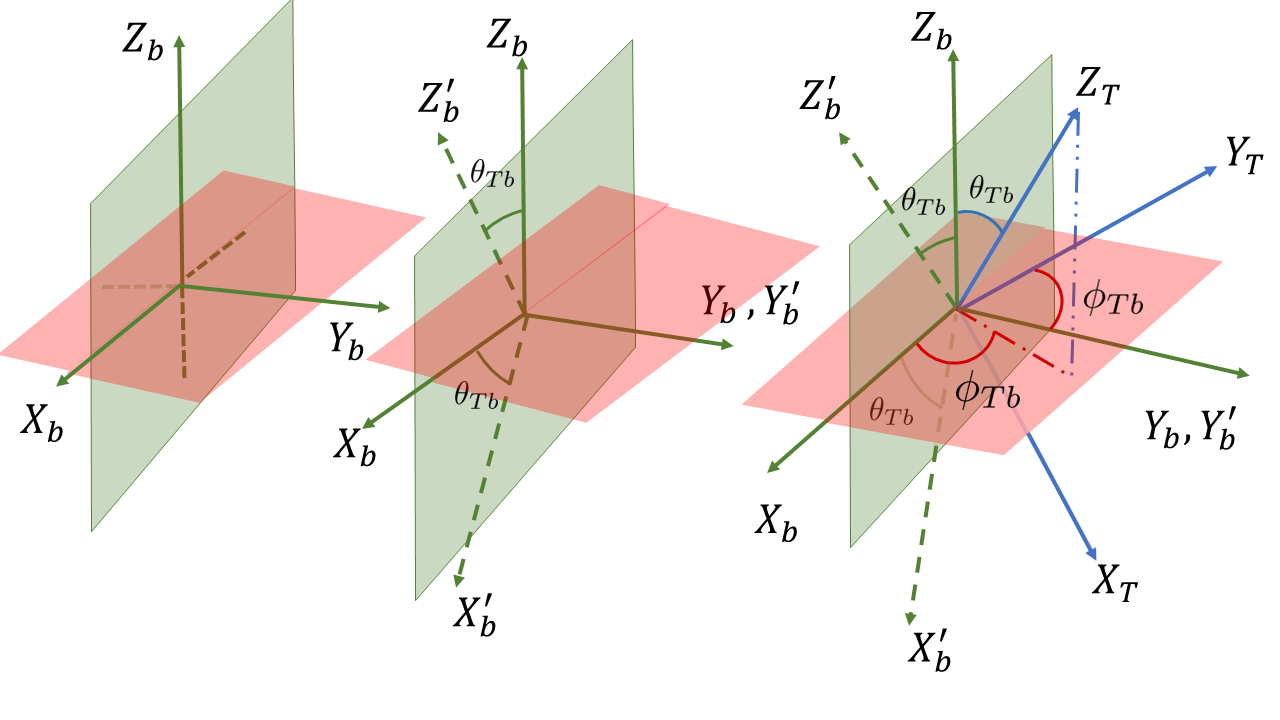}
	\caption{The transformation from the b-frame to the T-frame. The green plane is the $ Z_b X_b$ plane and the red plane is the $ X_b Y_b$ plane or the orbital plane.}
	\label{fig:Transformation_from_b_to_T}
\end{figure}

In Subsection \ref{subsubsect:T-frame}, we defined $\theta_{ Tb}$ as the angle between the ${ Z_T}$-axis and the ${  Z_b}$-axis and $\phi_{Tb}$ as the angle between the ${ Y_T}$-axis and the ${ Y_b}$-axis. Then the transformation from the b-frame to the T-frame can be done in two steps as:

\begin{enumerate}

\item A rotation of the b-frame about the $Y_b$-axis by an angle $\theta_{ Tb}$ gives a new frame ${ X^\prime_b} { Y^\prime_b} { Z^\prime_b}$. This transformation is shown in the middle panel of Fig. \ref{fig:Transformation_from_b_to_T}.

Here, the $ Y^\prime_b$-axis is along the $ Y_b$-axis and both of the $ Z^\prime_b$-axis and the $ X^\prime_b$-axis are in the $ Z_b X_b$ plane. The angle between the $ X_b$-axis and the $ X^\prime_b$-axis is the same as the angle between the $Z_b$-axis and the $ Z^\prime_b$-axis, i.e., $\theta_{ Tb}$.

\item Next, a rotation of the ${ X^\prime_b} { Y^\prime_b} { Z^\prime_b}$ frame about the $Z_b$-axis by an angle $\phi_{Tb}$ gives the $ X_T Y_T Z_T$ frame. This transformation is shown in the right panel of Fig. \ref{fig:Transformation_from_b_to_T}.

\end{enumerate}

The above two transformations can be written as:
\begin{equation}
	{ X_b \, Y_b \, Z_b} \xrightarrow[]{R^P_{ Y_b}(\theta_{  Tb})}  { X_b^{\prime} \, Y_b^{\prime} \, Z_b^{\prime} } \xrightarrow[]{R^P_{ Z_b}(\phi_{ Tb})}  { X_T \, Y_T \, Z_T} ~.
	\label{eq:btrans1}
\end{equation}

Hence, the reverse transformation can be written as:
\begin{equation}
	{ X_T \, Y_T \, Z_T} \xrightarrow[]{R^P_{ Z_b}(-\phi_{ Tb})}  { X_b^{\prime} \, Y_b^{\prime} \, Z_b^{\prime} } \xrightarrow[]{R^P_{ Y_b}(-\theta_{ Tb})}  { X_b \, Y_b \, Z_b} ~.
	\label{eq:btrans2}
\end{equation}

Thus, if a vector is denoted by $\overrightarrow{V}_{T}$ in the T-frame and by $\overrightarrow{V}_{ b}$ in the b-frame, these two would be related by
\begin{equation}
	\overrightarrow{V}_{ b}  = R^A_{ Z}(\phi_{ Tb}) \,  R^A_{ Y}(\theta_{Tb}) \overrightarrow{V}_{ T} ~.
	\label{eq:Ttransb}
\end{equation}  

Note that, if to the b-frame, one first applies a rotation about the ${ X_b}$-axis by an angle $\theta_{ Tb}$ followed by a rotation about the $ Z_b$-axis by an angle $\phi_{ Tb}$, then also one would obtain a new frame whose $Z$-axis would make an angle $\theta_{ Tb}$ with the ${  Z_b}$-axis and the $X$-axis would make an angle $\phi_{  Tb}$ with the ${X_b}$-axis. However, in that case, the ${  Y}$-axis of the new frame would not be in the orbital plane, which would contradict our choice of the ${  Y_T}$-axis being along $\widehat{r}_{  b}$.

\subsection{Transformation from the sc-frame to the sc$^\prime$-frame}
\label{sc_to_scprime}

The sc-frame is defined with its origin at the center of the spinning companion. We shift this origin to the barycenter keeping the directions of the axes unchanged. This gives a new frame called the sc$^\prime$-frame.

As described in Subsection \ref{subsubsect:scprime-frame}, that the sc-frame (origin at the center of the companion) and the sc$^\prime$-frame (origin at the center of the companion) are related by a translation without any rotation, any vector in these two frames is basically the same, i.e., if a vector in the sc-frame be denoted by $\vec{V}_{sc}$ and the same vector in the sc$^\prime$-frame by $\vec{V}_{sc^\prime}$, we can write, 
\begin{equation}
\label{eq:compscVecscprime}
	\vec{V}_{sc} = \vec{V}_{sc^\prime}.
\end{equation}

\subsection{Transformation from the the ${sc}^{\prime}$-frame to the s-frame}
\label{sc_to_s}

The transformation from the ${sc}^{\prime}$-frame to the s-frame can be understood from the definition of the ${sc}^{\prime}$-frame as described in Subsection \ref{subsubsect:scprime-frame}. This can be done in two steps as described below.

\begin{figure}[h]
	\includegraphics[width=150mm]{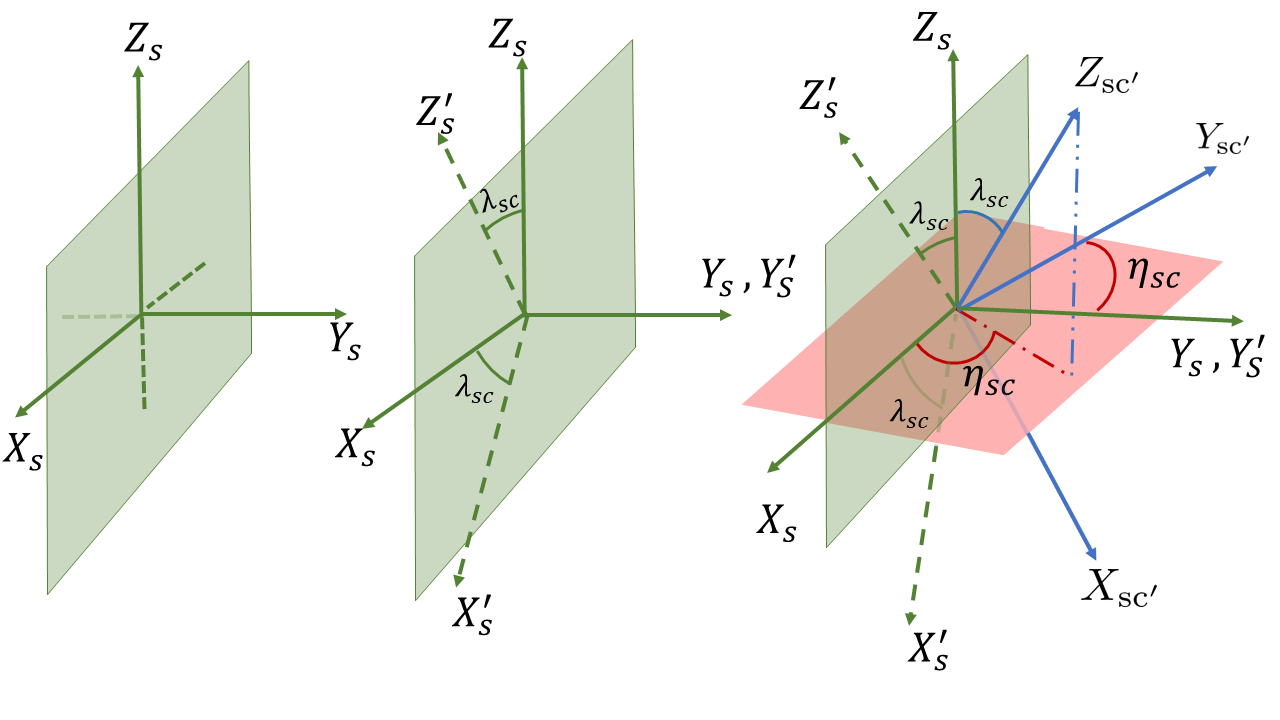}
	\caption{The transformation from the s-frame to the $  sc^\prime$-frame. The green plane is the $  Z_s X_s$ plane and the red plane is the $  X_s Y_s$ plane.}
	\label{fig:Transformation_from_S_to_bh}
\end{figure}

\begin{enumerate}

\item A rotation of the s-frame about the ${Y_s}$-axis by an angle $\lambda_{sc}$ gives a new frame ${  X_s^{\prime} \, Y_s^{\prime} \, Z_s^{\prime}}$ where the ${  Y_s^{\prime}}$-axis is along the ${  Y_s}$-axis, i.e., the ${  Y_s^{\prime}}$-axis is sill in the sky plane (the ${  X_s \, Y_s}$ plane). This transformation is shown in the middle panel of Fig. \ref{fig:Transformation_from_S_to_bh}. Here, we see that the ${  Z_s^{\prime} }$-axis and the ${  X_s^{\prime} }$-axis are in the $  Z_s \,   X_s$ plane. The angle between the $Z_s$-axis and the $  Z_s^{\prime}$-axis as well as the angle between the $  X_s$-axis and the $  X_s^{\prime} $-axis is $\lambda_{  sc}$.

\item Next, a rotation of the ${  X_s^{\prime} \, Y_s^{\prime} \, Z_s^{\prime}}$ frame about the ${Z_s}$-axis by an angle $\eta_{sc}$ gives the ${  X^\prime_{  sc} \, Y^\prime_{  sc} \, Z^\prime_{  sc}}$ frame. This transformation is shown in the right panel of Fig. \ref{fig:Transformation_from_S_to_bh}. As the rotation is about the ${  Z_s}$-axis, ${  Y^\prime_{  sc}}$ still remains in the ${  X_s \, Y_s}$ plane, i.e., in the sky-plane.

\end{enumerate}

The above set of transformations can be summarized as:
\begin{equation}
	{  X_s \, Y_s \, Z_s} \xrightarrow[]{R^P_{  Y_s}(\lambda_{  sc})}  {  X_s^{\prime} \, Y_s^{\prime} \, Z_s^{\prime} } \xrightarrow[]{R^P_{  Z_s}(\eta_{  sc})}  {  X^\prime_{  sc} \, Y^\prime_{  sc} \, Z^\prime_{  sc}} \, .
	\label{eq:ptrans1_bh_prime}
\end{equation}
Hence, the reverse transformation can be written as:
\begin{equation}
	{  X^\prime_{  sc} \, Y^\prime_{  sc} \, Z^\prime_{  sc}} \xrightarrow[]{R^P_{  Z_s}(-\eta_{  sc})}  {  X_s^{\prime} \, Y_s^{\prime} \, Z_s^{\prime} } \xrightarrow[]{R^P_{  Y_s}(-\lambda_{  sc})}  {  X_s \, Y_s \, Z_s} \, .
	\label{eq:ptrans2_bh_prime}
\end{equation}

Thus, if a vector is denoted by $\overrightarrow{V}_{  sc^\prime}$ in the sc$^{\prime}$-frame and by $\overrightarrow{V}_{s}$ in the s-frame, then these two would be related as follows:
\begin{equation}
	\overrightarrow{V}_{s}  = R^A_{  Z}(\eta_{  sc}) \,  R^A_{  Y}(\lambda_{sc}) \overrightarrow{V}_{sc^\prime} \, .
	\label{eq:atrans1_bh_prime}
\end{equation}

\subsection{Relating the forms of a vector in the sc-frame and in the b-frame}
\label{sub:vec_scprime_to_b}

Using the right hand side of Eq. (\ref{eq:atrans1_bh_prime}) in the left hand side of Eq. (\ref{eq:atrans_reverse2}), we get,
\begin{equation}
	R^A_{  Z}(\eta_{  sc}) \,  R^A_{  Y}(\lambda_{  sc}) \overrightarrow{V}_{  sc^\prime}  = R^A_{  X}(i)  R^A_{  Z}(\omega)  \overrightarrow{V}_{  b} ~.
\end{equation}
Hence, the relation between the forms of any given vector in the b-frame ($\overrightarrow{V}_{b}$) and in the sc$^{\prime}$-frame ($\overrightarrow{V}_{sc^\prime}$) is given by:
\begin{equation}
	\label{eq:btransbhpr}
	\overrightarrow{V}_{sc^\prime}  =R^A_{  Y}(-\lambda_{sc})\,  R^A_{  Z}(-\eta_{sc}) \, R^A_{  X}(i)  R^A_{Z}(\omega)  \overrightarrow{V}_{b} ~.
\end{equation}

Eq. (\ref{eq:compscVecscprime}) implies that the relation between the forms of any given vector in the b-frame ($\overrightarrow{V}_{b}$) and in the sc-frame ($\overrightarrow{V}_{sc}$) is the same, i.e., 
\begin{equation}
	\label{eq:btransbh}
	\overrightarrow{V}_{sc}  =R^A_{  Y}(-\lambda_{sc})\,  R^A_{  Z}(-\eta_{sc}) \, R^A_{  X}(i)  R^A_{Z}(\omega)  \overrightarrow{V}_{b} ~.
\end{equation}

\subsection{Relating the forms of a vector in the sc-frame and in the T-frame: Transformation from the ${  sc^\prime}$-frame to the T-frame}

Using the right hand side of Eq. (\ref{eq:Ttransb}) into the right hand side of Eq. (\ref{eq:btransbh}), we get the relation between the forms of any given vector in the sc$^{\prime}$-frame ($\overrightarrow{V}_{sc^{\prime}}$) and in the T-frame ($\overrightarrow{V}_{T}$) as:
\begin{equation}
	\label{Rel_vec_bhpr_T}
	\overrightarrow{V}_{  sc^\prime}  =R^A_{  Y}(-\lambda_{sc})\,  R^A_{  Z}(-\eta_{sc}) \, R^A_{X}(i)  R^A_{Z}(\omega)  R^A_{  Z}(\phi_{  Tb}) \,  R^A_{  Y}(\theta_{  Tb}) \overrightarrow{V}_{  T} ~.
\end{equation}

Eq. (\ref{eq:compscVecscprime}) implies that the relation between the forms of any given vector in the sc-frame ($\overrightarrow{V}_{sc}$) and in the T-frame ($\overrightarrow{V}_{T}$) is the same, i.e., 
\begin{equation}
	\label{Rel_vec_bh_T}
	\overrightarrow{V}_{  sc}  =R^A_{  Y}(-\lambda_{sc})\,  R^A_{Z}(-\eta_{sc}) \, R^A_{  X}(i)  R^A_{  Z}(\omega)  R^A_{  Z}(\phi_{  Tb}) \,  R^A_{  Y}(\theta_{  Tb}) \overrightarrow{V}_{T} ~.
\end{equation}

\subsection{Transformation from the L-frame to the T-frame}
\label{L_to_T}

We have defined the T-frame in Subsection \ref{subsubsect:T-frame} and the L-frame in Subsection \ref{subsubsect:L-frame} where we have seen that the $X_T$ axis makes an angle $- \left(\pi/2 - \psi_{  L,in}\right)$ with the $X_L$ axis (see Fig. \ref{fig:f3}). Hence, the T-frame can be obtained from the L-frame by rotating it about the ${  Z_L}$-axis through as:  
\begin{equation}
	{  X_L \, Y_L \, Z_L }
	\xrightarrow[]{R^P_{  Z_L}\left(-(\pi/2-\psi_{  L,in})\right)}
	{  X_T \, Y_T \, Z_T}.
\end{equation}

Thus, if a vector is denoted by $\overrightarrow{V}_{  L}$ in the L-frame and by $\overrightarrow{V}_{  T}$ in the T-frame,  
the two representations are related by  
\begin{equation}
	\overrightarrow{V}_{  T} = 
	R^A_{  Z}\left(\pi/2-\psi_{  L,in}\right) \overrightarrow{V}_{  L}.
	\label{eq:TtransL}
\end{equation}

\section{Discussions}

In this short review, we have synthesized and explicitly defined the primary coordinate frames commonly used to study astrophysical phenomena around pulsars, along with the mathematical relations required to transform between them. We hope these explicit definitions will serve as a clear, practical reference for analytical studies across pulsar physics - whether applying post-Newtonian approximations in binary systems, modeling the general relativistic effects around a strongly gravitating object (like a black hole), or mapping the local magnetospheric geometry of pulsars.

As the Square Kilometre Array Observatory (SKAO) telescopes come online, high sensitivity of the array and its broad radio frequency coverage will transform pulsar astronomy. For instance, the discovery of exotic pulsars, particularly near the Galactic center \citep{Abbate2025}, will offer unprecedented opportunities to test gravity in strong-field regimes \citep{vivek2025}. Interpreting such systems requires precise frame conversions between the observer's sky plane, orbital axes, and pulsar I-frame. 

Similarly, Pulsar Timing Array (PTA) science with the SKAO telescopes aims to detect nanohertz gravitational waves and conduct fundamental tests of physics \citep{Shannon2025}. Consistent coordinate definitions across local pulsar frames, Solar System barycentric coordinates, and observational frames are essential to eliminate alignment errors in high-precision timing models. Furthermore, the unprecedented sensitivity of the SKAO telescopes will allow us to probe emission mechanics and magnetospheric structures in much greater detail \citep{Oswald2025}, the modeling of which requires seamless transformations between the corotating magnetic dipole m-frame and the pulsar I-frame. 

Finally, future discoveries and higher observational precision will likely reveal subtle relativistic features, such as how light bending in curved spacetime affects pulse polarization profiles (Debnath et al., in prep.). Correctly interpreting these signatures requires robust analytical models grounded in clear coordinate frame transformations. By establishing a unified convention for these reference frames, we hope this review provides a reliable mathematical foundation for the analytical study and modeling of pulsar physics.

\end{document}